\documentclass[epj,referee]{svjour}
\begin{document}
\title{Magnetic Thomas-Fermi-Weizs\"acker model for quantum 
dots: a comparison with Kohn-Sham ground states} 
\titlerunning{Magnetic Thomas-Fermi-Weizs\"acker model for quantum 
dots \ldots} 
\author{Lloren\c{c} Serra \and Antonio Puente}
\authorrunning{Ll. Serra \and A. Puente}
\institute{Departament de F\'{\i}sica,
Universitat de les Illes Balears, E-07071 Palma de Mallorca, Spain}
\date{August 29, 2000}
\abstract{
The magnetic extension of the Thomas-Fermi-Weizs\"acker kinetic energy
is used within density-functional-theory to numerically obtain  
the ground state densities and energies of two-dimensional quantum dots. 
The results are thoroughly compared with the microscopic Kohn-Sham ones
in order to assess the validity of the semiclassical method. Circular 
as well as deformed systems are considered.
} 
\maketitle

\section{Introduction}

Semiclassical approaches to many-body systems are a very valuable 
tool since they provide physical insights which otherwise are very 
difficult to achieve. In fact, they have been applied since many years
ago to describe different systems such as atoms, atomic nuclei, metals 
and, more recently, metallic clusters and electronic nanostructures.
Two dimensional quantum dots are not an exception and have been 
analyzed using the Thomas-Fermi models in, for instance,  
Refs.\ \cite{Mce92,Mar92,Lie95,Bar98,Zar96,Pin00}. 

In Ref.\ \cite{Pue00} we performed calculations for quantum dots
using a selfconsistent Thomas-Fermi-Weizs\"acker (TFW) model similar to 
that developed by Zaremba and coworkers \cite{Zar96}. 
It is our aim in this paper to extend those calculations by including 
the effect of a perpendicular magnetic field $B$ using the magnetic 
extension of the kinetic energy within density-functional theory. 
As in Ref.\ \cite{Pue00} we will pay special attention to the 
quantitative comparison with the microscopic Kohn-Sham solution in order
to asses the accuracy limits of the TFW densities and energies as a 
function of $B$.  
The magnetic extension of the Thomas-Fermi theory was rigorously 
presented by Lieb {\em et al.} \cite{Lie95} and a selfconsistent 
numerical application to circular dots was done in Ref.\ 
\cite{Bar98} but, to our knowledge, no detailed 
comparison with microscopic calculations has been given in the
literature. We also present in this manuscript symmetry 
unrestricted calculations for deformed dots that had not been 
considered before within this model.

A peculiar characteristic of the magnetic Thomas-Fermi functional 
is its first-derivative discontinuity in the density 
dependence \cite{Lie95}, at particular density values. The physical
origin for this is found in the formation of constant energy  
Landau-bands in the non-interacting Fermi gas at certain magnetic fields. 
This implies that the 
mean field is discontinuous, thus manifesting 
the formation of {\em incompressible regions} in the system, each 
one characterized by the number of full Landau bands, {\em i.e.}, the 
filling factor $\nu$. 
It is worth to point out that experimental evidences of incompressible 
stripes at the edges of quantum dots and antidots have been obtained 
by means of far-infrared spectroscopy \cite{Bol96}.
As we will show, the semiclassical Euler-Lagrange 
equations selfconsistently determine the density profile, energy 
and chemical potential of the quantum dot,
which turn out to be in an overall good
agreement with full Kohn-Sham results for increasing magnetic fields
up to $\nu=1$. 

Section 2 of the paper is devoted to the presentation of the energy 
functional as well as the minimization equations. In Sec. 3 the results 
for circular as well as deformed quantum dots are given and, finally, 
Sec.\ 4 presents the conclusions. 

\section{Magnetic TFW functional}

Using a local approximation to density functional theory we assume 
that the energy of the system can be written in terms of the electronic
spin densities $\rho_\eta({\bf r})$, where $\eta=\uparrow,\downarrow$,
as $E=\int{d{\bf r}\, {\cal E}[\rho_\uparrow,\rho_\downarrow;B]}$. 
Notice also the explicit dependence on the magnetic field $B$ which is 
considered as a functional parameter.
The different contributions to the energy density may be written as
\begin{eqnarray}
\label{eq1}
{\cal E}&\!\!\![&\!\!\! \rho_\uparrow,\rho_\downarrow;B]=
\tau[\rho_\uparrow;B]+
\tau[\rho_\downarrow;B] 
+{1 \over 2} v_H({\bf r}) \rho \nonumber\\
&+&{\cal E}_{XC}(\rho_\uparrow,\rho_\downarrow)+
v_{\rm ext}({\bf r})\rho
+{\cal E}_Z(\rho_\uparrow,\rho_\downarrow;B)
\; ,
\end{eqnarray}
where the first two pieces give the kinetic energies 
of spin up and down electrons; the third is the Hartree energy
in terms of the Hartree potential $v_H$ and the total density 
$\rho=\rho_\uparrow+\rho_\downarrow$; the fourth is the 
exchange-correlation contribution; the fifth is the 
energy due to the external potential $v_{\rm ext}$ and,  
the last one corresponds to the Zeeman contribution.
As in Refs.\ \cite{Zar96,Pue00},
the kinetic energy contains a pure Thomas-Fermi term and a 
gradient (Weizs\"acker) one, $\tau=\tau_{TF}+\tau_W$. The gradient 
term is given by
\begin{equation}
\label{eq2}
\tau_W[\rho_\eta]={\hbar^2\over 2m} \lambda 
{(\nabla\rho_\eta)^2\over\rho_\eta}\; ,
\end{equation}
with $\lambda=1/4$, while the Thomas-Fermi piece is \cite{Lie95,Bar98}
\begin{eqnarray}
\label{eq3}
\tau_{TF}[\rho_\eta;B] &=&
{1\over 2}\, \hbar\omega_c\, D\, S_\eta^2 \nonumber\\
&+& \hbar\omega_c\, \left(
S_\eta + {1\over 2}\right)\,
\left(\rho_\eta-S_\eta D\right)\; .
\end{eqnarray}
In this last expression $\omega_c={e B\over m c}$ is the cyclotron 
frequency, $D={eB\over 2\pi\hbar c}$ is the Landau level degeneracy 
per unit area and $S_\eta=[{\rho_\eta\over D}]$ is the integer part 
\cite{not0} of the local filling factor $\nu_\eta={\rho_\eta\over D}$ 
and gives the index
of the highest fully occupied Landau band. The two contributions in
Eq.\ (\ref{eq3}) give therefore the kinetic energy of the fully 
occupied Landau bands and that corresponding to the last partially filled band, 
respectively. It can be shown \cite{Lie95} that, in the limit 
$B\to 0$,   the non-magnetic Thomas-Fermi 
functional $\tau_{TF}(\rho,B=0)={\hbar^2\over 2m}\pi\rho^2$ is 
recovered from Eq.\ (\ref{eq3}).

For the exchange-correlation energy ${\cal E}_{XC}$ we have used the 
LSDA functional based on the Tanatar-Ceperley calculations for 
the 2D uniform electron gas \cite{Tan89} and the von Barth-Hedin 
interpolation for intermediate polarizations \cite{Bar72}. The expression
can be found, {\it e.g.}, in Ref.\ \cite{Kos97}, where Kohn-Sham
results for parabolic dots at $B=0$ were given. Notice that current-density 
dependence is not included in the functional. This could in principle be 
done within the so-called current-density-functional theory although it 
is known that the contribution to the ground state energy from these terms 
is in general quite small and only at very high magnetic fields they can 
be of relevance \cite{Fer94,Rei99}. 

The Zeeman energy 
reads, in terms of the effective gyromagnetic factor $g^*$ and
Bohr magneton $\mu_B$, 
\begin{equation}
\label{eq4}
{\cal E}_Z(\rho_\uparrow,\rho_\downarrow;B)=
{1\over 2}\, g^*\, \mu_B\, B\, (\rho_\uparrow-\rho_\downarrow)\; .
\end{equation}

The energy functional is minimized by the ground state spin densities,
or equivalently by the ground state total density $\rho$ and 
magnetization $m=\rho_\uparrow-\rho_\downarrow$, with the constraint
of conservation of the total number of particles. The 
corresponding Lagrange parameter $\mu$ is, by definition, the chemical 
potential. The two sets of equivalent equations read
\begin{equation}
\label{eq5}
\left\{\begin{array}{c}
{\delta E\over \delta\rho}=\mu\\[0.2cm]
{\delta E\over \delta m}=0
\end{array}\right.\qquad
\Leftrightarrow\qquad
\left\{\begin{array}{c}
{\delta E\over \delta\rho_\uparrow}=\mu\\[0.2cm]
{\delta E\over \delta\rho_\downarrow}=\mu
\end{array}\right.\; .
\end{equation}
For convenience, we choose to work with the second set  
which can be transformed, introducing new variables 
$\psi_\eta=\sqrt{\rho_\eta}$, into the alternative Schr\"odinger-like 
equations 
\begin{eqnarray}
\label{eq6}
-4\lambda{\hbar^2\over 2m} \nabla^2\psi_\eta &+& 
\left( v_{\rm ext}+v_H+
{\partial{\cal E}_{XC}\over\partial\rho_\eta} \right.\nonumber\\
&+& \left. C_\eta + \alpha_\eta {g^*\over 2} \mu_B B
\right)
\psi_\eta = 
\mu \psi_\eta\; ,
\end{eqnarray}
where we have defined $\alpha_\uparrow=1$, $\alpha_\downarrow=-1$
and also introduced the contribution from the Thomas-Fermi energy
\begin{equation}
\label{eq7}
C_\eta=\hbar \omega_c \left( S_\eta+{1\over 2} \right)\; .
\end{equation}

The solution of the two coupled Eqs.\ (\ref{eq6}), {\it i.e.}, for each spin 
component, has been obtained numerically by discretizing the two dimensional 
$xy$ plane into a uniform grid of points and using the imaginary 
time-step method. The grid size is typically $70\times 70$ points, while 
the Laplacian operator is discretized by using 7 points formulas. The 
stability of the results when increasing these values has been ckecked. 
The Kohn-Sham results, we will compare with below, have been 
obtained using a similar method developed by us in Ref.\ \cite{Pue99}. 

\section{Results}

\subsection{The TF plateaus}
We begin the results section by discussing the effect of the 
discontinuous contribution to the potential.  
We present here calculations for two different dots: a circular one 
containing $N=42$ electrons under parabolic confinement 
$v_{\rm ext}=1/2 \, \omega_0^2 \, r^2$ \cite{units,not1} and a 
 second one containing $N=20$ electrons in a deformed parabola
\begin{equation}
v_{\rm ext}({\bf r}) = {1\over 2} \omega_0^2 {4\over (1+\beta)^2}
(x^2+\beta^2 y^2)\; ,
\end{equation}
with anisotropy factor \cite{not2} $\beta=0.75$ and a coefficient
$\omega_0$ given by $N_p=20$ and the same $r_s$ as for $N=42$.

The spin up and down densities for the circular dot are displayed 
in the upper-left panel of Fig.\ \ref{fig1}, in comparison with the corresponding 
Kohn-Sham ones. The horizontal 
dotted lines indicate the values $\rho=D$ and $2D$ where $D$ is the 
density for bulk filling factor $\nu=1$ (Sec.\ 2). 
The TFW density is clearly 
giving flat regions, {\it i.e.}, plateaus, at the densities of the bulk
integer filling factors, which must be associated with incompressible
stripes in the finite system. Due to the effect of the Weizs\"acker
term the transition between different plateaus is quite smooth.  
In this particular case the spin down density has attained the 
first plateau and is beginning to fill a second one
around the dot center. On the other hand, the spin up density has 
already reached the second plateau at the center of the dot. 

The correlation 
of the plateaus with Landau bands is made even clearer in the upper-right 
panel, where the Kohn-Sham eigenvalues are plotted as a function of 
orbital angular momentum $\ell$. In this plot the horizontal line at 
$\epsilon \approx -1.24 \, {\rm H}^*$ indicates the Fermi energy. 
A proportionality between $\ell$ and $r$ may be 
established by noting that high $\ell$ values imply outer orbits.
Therefore, at the dot center (low $\ell$'s) two spin up bands are
filled while for spin down the second band is only partly occuppied.
When going towards the edge (increasing $\ell$) the second  
spin down band is rapidly depleted and at a larger $r$ the same happens
with the second spin up band. This behaviour of the microscopic solution
is in excellent agreement with that inferred from the plateaus of the
left panel, thereby showing the quality of the model. The 
formation of the plateaus is not as clear in the KS densities  
because of rather large density oscillations, quite similar to 
the Friedel oscillations found in metals. 

The two lower plots of Fig.\ \ref{fig1} show the plateaus in spin up and down 
densities for the deformed dot at $B=1$ T. A behaviour similar to that 
of the circular case is inferred, although in this case the plateaus
adjust their shape to the anisotropy $\beta=0.75$ of the confining 
potential. At larger magnetic fields, however, deviations are obtained 
as we will show when presenting the systematic results in Sec.\ 3.3

\subsection{Circular dot with 42 electrons}
In this subsection we show in a systematic way the results for the dot 
containing $N=42$ electrons in a circular parabola, with 
$r_s=1.5\; a_0^*$ and $N_p=42$.
Figure \ref{fig2} displays the evolution 
with magnetic field of the density and magnetization profiles 
in comparison with the Kohn-Sham ones. 
In general the TFW density and magnetization are correctly averaging the KS 
values with a rather good agreement for all the magnetic fields considered.  
At $B=5$~T the TFW correctly yields equal density and 
magnetization distributions, due to the achievement of full 
polarization. At this magnetic field 
the KS result corresponds to the maximum-density-droplet (MDD) 
solution \cite{MDD}, in which the single-particle angular momenta are 
succesively occupied up to the $\ell_{\rm max}=N-1$ value. 
Increasing the magnetic field still further, {\it i.e.}, entering the region 
of fractional filling 
factor $\nu<1$, the dot evolves by reconstructing the edge, as seen 
in the $B=6$ and 7~T panels. However, this physical behaviour is 
apparently not well reproduced by the TFW model.  

In the upper-left panel of Fig.\ \ref{fig3} a quantitative comparison between the 
energy per particle in TFW and KS is provided. We see that the 
difference remains below $2$\%, 
although on the plot it may seem magnified because of the expanded 
scale. In the upper-right panel a comparison of the TFW chemical potential
$\mu$ with the KS Fermi energy is given, which is again indicating
the good estimate given by the TFW of the less bound electron, until the 
edge reconstruction begins. We mention that we remain here in the limit 
$T\to 0$, although a small value of $T$ is sometimes necessary to converge 
the KS results.

\subsection{Elliptic dot with $N=20$}
The lower panels in Fig.\ \ref{fig3} represent the comparison of the ground state 
energy per particle $E/N$ and TFW chemical potential {\it vs} KS Fermi energy 
corresponding to the 20-electron dot in a deformed parabola (Sec.\ 3.1). 
As for the previous circular system the agreement between the TFW and KS ground 
state energies is remarkable, not exceeding in this case a $3$\%.

Figures \ref{fig4} and \ref{fig5} show within TFW and KS, respectively, the local  
filling factors $\nu$ for selected values of the magnetic field. 
A similar behaviour as that discussed for the circular case is obtained. 
In fact we recognize the preference of the TFW density to produce 
flat regions associated with the integer filling factors. By looking 
at the central region we can also identify the progressive 
depletion of the central plateau when increasing the 
magnetic field. For instance, one can follow the evolution $\nu=3,2,1$ 
for $B=1,1.5,2.5$ T, respectively. This behavior is also inferred from the KS
results (Fig.\ \ref{fig5}) although it is somehow masked by the large oscillations
of quantum origin. As in the circular case, the prediction of the dot 
polarization with magnetic field is also in good agreement with the microscopic 
result. This can be qualitatively seen in Figs.\ \ref{fig4} and \ref{fig5} 
by noting the clear 
depletion of the spin down density starting at $B=3$ T which ends with a fully 
polarized dot ($S=10$) for $B \ge 4.5$ T. 

A conspicuous prediction of the KS model is the gradual change in shape 
of the quantum dot when increasing the magnetic field. This is quite 
evident for $B\ge 4$~T, when the dot can no longer be considered 
elliptic, but rather rectangular in shape. We attribute this to the
competition between the symmetry of the external field and the 
preference for circular edge reconstructions induced by large magnetic 
fields. A more abrupt transition to a circular shape was obtained 
in Ref.\ \cite{UJM} within the ultimate jellium model, which permits 
the deformation of the external potential. In our model this is fixed
and, therefore, it seems natural that a stronger competition is 
present. When comparing with the TFW results, we notice that although
the central plateau indeed seems to evolve towards a rectangle 
(for $B=4.5-5.5$ T), the 
outer edge remains always elliptic. The defficiency of the semiclassical 
model in reproducing morphological changes attributed to the magnetic 
field can be explained by the fact that within TFW the magnetic field effects
are taken into account in a purely local way (see  Eq.\ (\ref{eq7})),
and thus can hardly induce any influence on the global shape.
Another missing feature in the TFW results is the incipient electron 
localization seen in the KS panels for $B>6$~T (Fig.\ 5). 
This localization has also been predicted within Hartree-Fock 
theory \cite{Mul96} and more recently,
using current-density-functional theory \cite{Rei99}.

\section{Conclusions}

The validity limits of the semiclassical TFW model have been 
discussed by comparing with the microscopic Kohn-Sham model.
The magnetic extension of the semiclasical kinetic energy
produces a discontinuous mean field which favours the appearance
of density plateaus corresponding to integer filling factors
for the bulk gas. The correspondence of these plateaus with 
the occupation of Landau bands in the finite systems has been 
proved for a circular dot with $N=42$ electrons. In circular dots, 
the systematic evolution with $B$ of the density and magnetization
profiles is rather well predicted by the TFW model up to filling factor 
$\nu=1$. The same happens with the dot energy and chemical potential. 
In deformed dots the ground state energy is also well reproduced by TFW, 
although this model is not able to yield the shape changes induced
by the magnetic field, as found in the KS result.      

\vspace{0.5truecm}

This work has been performed under Grant No.\ PB98-0124 from DGESeIC, Spain.

\begin{figure}[!htb]
\caption{Upper row: Left panel shows spin up and down densities with 
TFW (solid) and Kohn-Sham (dashed) models. Dotted lines indicate the 
bulk densities for filling factor 1 and 2. Right panel displays the 
(spin up/down) KS eigenvalues (up/down triangles) as a function of orbital 
angular momentum. The horizontal line indicates the Fermi energy 
(effective atomic units  \cite{units}
are used). Lower row: spin densities for 
a deformed quantum dot (see Sec.\ 3).
}
\label{fig1}
\end{figure}
\begin{figure}[!bth]
\caption{Evolution with magnetic field of density $\rho (r)$ and magnetization 
$m(r)$ for the circular dot with $N=42$. Solid lines correspond to TFW 
and dashed ones to KS. In each panel the curves with lower values 
at the origin correspond to $m(r)$, except for $B>4$ T where 
the dot is fully polarized and thus $m=\rho$.}
\label{fig2}
\end{figure}
\begin{figure}[!tbh]
\caption{Upper panels: (left) energy per particle within the TFW (solid) and KS 
(dots) models; (right) TFW chemical potential (solid) {\it vs} KS Fermi energy 
(dots) for the circular system with $N=42$ electrons. Lower panels: 
corresponding values for $N=20$ electrons in a deformed parabola.}
\label{fig3}
\end{figure}
\begin{figure}[!thb]
\caption{Evolution, within the TFW model, of the local filling factor with 
magnetic field for the elliptic dot with $N=20$ electrons. In each case 
spin up (left) and down (right) values are shown, except for $B \ge 4.5$ T 
in wich only the spin up result is represented since full polarization has 
been attained. White areas 
correspond to near-integer values (plateaus) while black contours indicate 
half-integer transition values.}
\label{fig4}
\end{figure}
\begin{figure}[!tbh]
\caption{Same as Fig.\ \ref{fig4} within KS. The ground state total spin is 
indicated 
at the bottom of the plot for each magnetic field. for $B \ge 4.5$ T, the dot 
is fully polarized and thus $S=10$.}
\label{fig5}
\end{figure}

\begin{thebibliography}{}
\bibitem{not0} For computational stability the integer-part
function $[x]$ is smoothed in the following way 
$[x]\approx \sum_{i=1}[1+\tanh({x-i\over\lambda})]/2$ with 
$\lambda=0.05 \; a_0^*$. For $\lambda$ small enough the results are
not sensitive on its precise value.
\bibitem{units} We use the standard effective units system,
defined by $\hbar=e^2/\epsilon=m=1$, where $\epsilon$ is the dielectric 
constant and $m$ is the effective electron mass (given in terms of the 
bare mass $m_e$ as $m=m^* m_e$). Accordingly, the length unit is given by 
the effective Bohr radius $a_0^*=\hbar^2\epsilon/me^2$ and the energy unit by 
the effective Hartree ${\rm H}^*=e^2/(2\epsilon a_0^*)$. 
Taking the GaAs values $\epsilon=12.4$ 
and $m^*=0.067$ we get H$^*\approx12$~meV and $a_0^*\approx 98$~\AA.
\bibitem{not1} The parabola constant is determined by 
$\omega_0^2=1/(\sqrt{N_p}r_s^3)$, with $N_p=42$ and $r_s=1.5 a_0^*$
 This definition of $\omega_0$ is used in order to 
reproduce the curvature at the origin of the potential of a jellium
disk containing $N_p$ positive charges and a radius per unit charge
given by $r_s$.
\end{thebibliography}
\end{document}